\documentclass[useAMS]{mn2e}
\usepackage{amsmath}
\usepackage{textcomp}
\usepackage{amssymb}
\usepackage[pdftex]{graphicx}
\usepackage{amssymb}

\usepackage[margin=.8in, paperwidth=8.5in, paperheight=11in]{geometry}
\usepackage{multirow}
\usepackage{multicol}
\usepackage{epsfig}
\usepackage{graphicx}
\usepackage{esint}
\usepackage{rotating}
\usepackage{pdflscape}
\usepackage{adjustbox}
\begin{document}

\title{Effects of Opacity Temperature Dependence on Radiatively Accelerated Clouds}
\author[S. Dyda et al.]
{\parbox{\textwidth}{Sergei~Dyda$^1$\thanks{sdyda@ast.cam.ac.uk}, Daniel~Proga$^2$, Christopher S. Reynolds$^1$}\\
$^{1}$Institute of Astronomy, Madingley Road, Cambridge CB3 0HA, UK \\
$^{2}$Department of Physics \& Astronomy, University of Nevada Las Vegas , Las Vegas, NV 89154
}

\date{\today}
\pagerange{\pageref{firstpage}--\pageref{lastpage}}
\pubyear{2019}

\label{firstpage}

\maketitle

\begin{abstract}
We study how different opacity-temperature scalings affect the dynamical evolution of irradiated gas clouds using time-dependent, radiation-hydrodynamics (rad-HD) simulations. When clouds are optically thick, the bright side heats up and expands, accelerating the cloud via the rocket effect. Clouds that become more optically thick as they heat accelerate $\sim 35\%$ faster than clouds that become optically thin. An enhancement of $\sim 85\%$ in the acceleration can be achieved by having a broken powerlaw opacity profile, which allows the evaporating gas driving the cloud to become optically thin and not attenuate the driving radiation flux. We find that up to $\sim 2\%$ of incident radiation is re-emitted by accelerating clouds, which we estimate as the contribution of a single accelerating cloud to an emission or absorption line. Re-emission is suppressed by ``bumps'' in the opacity-temperature relation since these decrease the opacity of the hot, evaporating gas, primarily responsible for the re-radiation. If clouds are optically thin, they heat nearly uniformly, expand and form shocks. This triggers the Richtmyer-Meshkov instability, leading to cloud disruption and dissipation on thermal time-scales.    
\end{abstract}


\section{Introduction}
Gas clouds appear in many astrophysical systems such as the interstellar medium (ISM), integalactic medium (IGM) and active galactic nuclei (AGN). In AGN, X-ray studies show that observed column densities are variable, suggesting that the torus and broad line region is clumpy and very dynamic (e.g. Krumpe, Markowitz \& Nikutta 2014, Ramos Almeida \& Ricci 2017 and references therein). Likewise multi-wavelength observations of galactic winds from radio to X-ray find multi-component, multi-temperature outflows of gas and dust (e.g Veilleux et al. 2005 and references therein). In some contexts (Tombesi et al. 2015), an important avenue of study is what physical processes are responsible for accelerating the clouds? In other cases, the important question is how efficiently is energy deposited from the radiation field to the gas?  

The key to modeling these systems is understanding the radiation and gas coupling. When the gas is mostly neutral, the coupling is dominated by bound-free interactions, whereas when it is ionized the interactions are dominated by free-free processes. Photoionization calculations (see for example Iglesias \& Rogers 1996) show that opacity is a complex function of gas density and temperature.  Before we introduce sophisticated microphysics into our models, such as Compton heating/cooling, radiation pressure due to spectral lines or dust we aim to understand how different opacity scalings as a function of temperature affect cloud dynamics.

We consider an opacity parametrized by a power-law 
\begin{equation}
\kappa = \kappa_0 \left( \frac{T}{T_0}\right)^{-s} \left( \frac{\rho}{\rho_0}\right)^{n-1},
\label{eq:opacity} 
\end{equation}
where $T$ and $\rho$ are the temperature and density respectively and the 0 subscript denotes some fiducial value. The powerlaw indices $s$ and $n$ encapsulate the microphysics responsible for the gas opacity.

Previous numerical simulations have studied clouds in different physical regimes. For example, Proga et al. (2014, hereafter P14) used the Kramers form of opacity, $s = 3.5$ and  $n = 2$ to study cloud evolution in the broad line region of AGN. They found the clouds, which are optically thick to absorption, disperse before they can move more than a few cloud radii. Zhang et al. (2018, hereafter Z18) studied the evolution of dusty clouds in rapidly star-forming galaxies where radiation flux is dominated by the IR. They used an opacity scaling with $s = -2$ and  $n = 1$ and found that clouds can be significantly accelerated without being dispersed. The scaling of opacity with temperature in these two studies is different, leading to qualitatively different responses from the irradiated cloud. In P14, as the cloud absorbs radiation, it heats up and because $s > 0$ becomes \emph{less} optically thick, thereby slowing down the heating.  In contrast, the Z18 case has $s < 0$, so cloud heating is a runaway process as opacity \emph{increases} with temperature.

These earlier cloud models used a simplified, power-law expression for the opacity (\ref{eq:opacity}). Photoionization calculations have shown that the Roseland mean opacity is not monotonic, with features due to H, He and Fe. The iron opacity peak has been shown to be important in the structure and stability of massive star envelopes (Jiang et al. 2015) and AGN discs (Jiang, Davis \& Stone 2016). On either side of these features the opacity scaling changes sign, potentially affecting the cloud dynamics. To build an intuition for models where opacity is computed self consistently with photoionization codes, we study cloud acceleration models where opacity scales like (\ref{eq:opacity}) for different temperature power-law scalings $s$.

Our models consist of over-dense, cold, spherical clouds in pressure equilibrium with a dilute, hot, ambient gas irradiated from one side. We consider two sets of simulations exploring the effects of the temperature scaling of cloud opacity. In one set of models, we keep the optical depth of the cloud constant but vary $\kappa_0$. In another set of models we keep the opacity of the ambient gas fixed, and vary the optical depth of the cloud. In all our models the cloud is initially optically thick and the ambient gas is optically thin.

We find two types of behaviour: clouds can \emph{balloon} outward or they may \emph{accelerate} away from the radiation source. The former occurs if the cloud becomes optically thin and thus heats nearly uniformally, as in the P14 models. The later occurs if the cloud remains optically thick and heats non-uniformally, accelerating the cloud via the rocket effect (Oort \& Spitzer 1955, hereafter OS55) as hot gas evaporates away, as in the Z18 models (see also Mellema 1998 and references therein). Based on the intuition developed from these power-law opacity models, we then consider models where the opacity scaling with temperature changes sign at a critical temperature, to model the effect of a ``bump'' in the opacity (see for example Fig 5.2 in Hansen et al. 2004). We find that this can change the heating rate or acceleration efficiency, but that it does not qualitatively change the dynamics. 

The outline of our paper is as follows. In Section \ref{sec:numerical}, we describe our numerical setup for modeling the clouds. In Section \ref{sec:monotonic}, we describe our main results for clouds with monotonic dependence of opacity on temperature and in Section \ref{sec:turnover} describe results for models with broken power-law opacity, simulating a feature in the opacity profile.  In Section \ref{sec:discussion}, we discuss applications of this work, in particular to modeling clouds in the broad line region of AGN and for heating gas in the IGM. We conclude in Section \ref{sec:conclusion} where we discuss the physical processes we would like to include in future simulations of clouds as well as the prospects for studying multi-cloud systems.

\section{Numerical Methods}
\label{sec:numerical}
We performed all numerical simulations with the developmental version of the radiation magnetohydrodynamics (rad-MHD) code \textsc{Athena++} (Stone et al. in prep), a re-write of the MHD code \textsc{Athena}  (Gardiner \& Stone 2005, 2008), optimized for adaptive mesh refinement and various modules incorporating new physics including, crucially for this work, radiation transport (Jiang, Stone \& Davis 2012, 2014). The basic physical setup is a 2D box with initially constant gas pressure, centered on an over-dense spherical cloud. Radiation flux enters the box along a fixed direction, which is assumed to be emitted from a far away blackbody, hotter than the gas. The radiation causes the cloud to heat, accelerate and shear, depending on the strength of the opacity. We describe our setup in more detail below.

\subsection{Basic Equations}
\label{sec:setup}
In dimensionless form the basic equations for single fluid hydrodynamics coupled to a radiation field are
\begin{subequations}
\begin{equation}
\frac{\partial \rho}{\partial t} + \nabla \cdot \left( \rho \mathbf{v} \right) = 0,
\end{equation}
\begin{equation}
\frac{\partial (\rho \mathbf{v})}{\partial t} + \nabla \cdot \left(\rho \mathbf{vv} + \mathsf{P} \right) = -\mathbb{P} \mathbf{S_r}(\mathbf{\mathbf{P}}),
\label{eq:momentum}
\end{equation}
\begin{equation}
\frac{\partial E}{\partial t} + \nabla \cdot \left( (E + P)\mathbf{v} \right) = - \mathbb{PC} S_r(E),
\label{eq:energy}
\end{equation}  
\label{eq:hydro}%
\end{subequations}
where $\rho$, $\mathbf{v}$ are the fluid density and velocity respectively and $\mathsf{P}$ is a diagonal tensor with components P the gas pressure. The total gas energy is $E = 1/2 \rho |\mathbf{v}|^2 + \mathcal{E}$ where $\mathcal{E} =  P/(\gamma -1)$ is the internal energy. The isothermal sound speed is $a^2 = P/\rho$ and the adiabatic sound speed $c_s^2 = \gamma a^2$. The temperature is $T = (\gamma -1)\mathcal{E}\mu m_{\rm{p}}/\rho k_{\rm{b}}$ where $\mu = 1.0$ is the mean molecular weight and other symbols have their standard meaning. The radiation source terms $\mathbf{S_r}$ and $S_r(E)$ are calculated from the difference between the angular quadratures of the specific intensity $I(\mathbf{n})$ along unit vecotrs $\mathbf{n}$ in the lab frame before and after adding the source terms (see Jiang, Stone \& Davis 2019). To provide some physical intuition, the radiation source terms are at lowest order in $v/c$ 
\begin{subequations}
\begin{align}
\begin{split}
\mathbf{S_r}(\mathbf{P}) &= - \left( \sigma_s + \sigma_a \right) \mathbf{F}_r  + \mathcal{O}\left( v/\mathbb{C}\right)
\end{split}
\label{eq:rad_mom}
\end{align}
\begin{align}
\begin{split}
S_r(E) &= \sigma_a \left(T^4 - E_r\right) + \mathcal{O}\left( v/\mathbb{C}\right),  
\end{split}
\end{align}
\end{subequations}
where the radiation energy density and flux are
\begin{subequations}
\begin{align}
E_r = 4 \pi \int I d\Omega,
\end{align}
\begin{align}
\mathbf{F}_r = 4 \pi c \int \mathbf{n} I d\Omega,
\end{align}
\end{subequations} 
and the absorption and scattering cross sections are $\sigma_a$ and $\sigma_s$ respectively and the integrals are over all solid angles $\Omega$.

\subsection{Initial Conditions}
Initially the box is in hydrostatic equilibrium with gas pressure $P_0$ and a circular cloud at the center of the grid with density profile
\begin{equation}
\rho = \rho_0 + \frac{\rho_1 - \rho_0}{1 + \exp(10(r-1))},
\end{equation}
where $\rho_1 = 200 \rho_0$ is the maximum cloud density and $r = (x/x_0)^2 + (y/y_0)^2$. Here $x_0 = y_0 = 0.05$ is the radius of the cloud. Because the higher density cloud is in pressure equilibrium with the ambient gas, its temperature is less than $T_0$.

Our setup is meant to simulate a cloud far from the radiation source. Using \textsc{Athena++} one can assume light is emitted isotropically for point sources, for rays along angles computed from the algorithm described in Lowrie et al (1999). Crucially, these rays are never parallel to coordinate axes, so to simulate plane parallel radiation we study cloud acceleration in a rotated coordinate system. We set our radiative sources to lie along the left and bottom parts of the box and have four, uniformly distributed rays in the 2D plane parallel to the diagonals of the box (i.e. we set the code parameter $n_{\rm{ang}} = 4$). This setup allows us to resolve the rectangularly shaped ``shadow" behind optically thick clouds irradiated by a plane parallel source. If only one side of the box is radiating, the shadow is not well resolved, being triangular in shape, becoming more rectangular as additional rays are used to trace the radiation. The price to pay for this setup is the cloud accelerates diagonally, so we describe all our results in the coordinate system rotated by $45^{\circ}$ in the clock-wise direction. All our analysis is performed in the rotated coordinate system where radiation enters from the left, $-x$ side of the box and the cloud accelerates to the right in the ($+x$) direction. We also measure radiation exiting the top $+y$ and bottom $-y$ of the box.     

The other issue is that for the radiation flux $\mathbf{F_r} = F_0 \hat{x}$ to correspond to a unidirectional beam of radiation emitted by a blackbody at temperature $T_{\rm{beam}}$, we require the temperature of the isotropically emitting boundary to be $T_{\rm{iso}} = 2^{5/8} T_{\rm{beam}}$. We want to simulate clouds radiatively accelerated by a blackbody at temperature $T_{\rm{rad}} = 2 T_0$, implying an intensity $I_{\rm{beam}} = 16 I_{0}$. Therefore the intensity at the left and lower boundaries are set to $I = 2^{5/2} I_{\rm{beam}} = 90.51 I_0$. These are more technical issues, and in the remainder of this paper we will refer to radiation of temperature $T_{\rm{rad}} = 2 T_0$. 

Radiation is coupled to the gas via an absorption coefficients $\sigma_a = \rho \kappa$. Substituting our parameterization for the opacity (\ref{eq:opacity}), the absortion coefficient is
\begin{equation}
\sigma_a = \sigma_{a,0} \left(\frac{T}{T_0}\right)^{-s} \left(\frac{\rho}{\rho_0}\right)^n.
\label{eq:absorption}
\end{equation}    
For simplicity, we set the scattering coefficient $\sigma_s = 0$. In our fiducial model the absorption coefficient $\sigma_{a,0} = 0.1$ and the optical depth of the cloud is initially $\tau_a = 2x_o \sigma_a = 2$. In one set of simulations we vary the power-law coefficient $s$ but keep $\sigma_{a,0} = \rm{const}$, thus effectively varying the cloud optical depth $\tau_a$. In the second set of models, as we vary $s$ we keep the optical depth $\tau_a = \rm{const.}$ and thus vary $\sigma_{a,0}$. In all cases the absorption optical depth of the ambient gas $\tau_{a,g} \leq 0.1$, so we refer to it as optically thin.

On the top and right sides of the box we impose outflow conditions on the gas variables and vacuum conditions on the radiation.  Along the bottom and left side of the box we keep density and pressure kept fixed at $\rho_0$ and $P_0$ respectively, while ensuring velocity is conserved when we perform this update.

Our simulation uses dimensionless parameters, but for AGN clouds reasonable parameters might be cloud temperature $T_0 = 2.44 \times 10^{6} \rm{K}$ and the cloud density is $\rho_0 = 1 \ \rm{g \ cm^{-3}}$. The pressure is then $P_0 = \rho_0 T_0 k_b/\mu m_p = 2.02 \times 10^{14} \rm{erg \ cm^{-3}}$, where we assumed $\mu = 1$. The isothermal sound speed $a_0 = \sqrt{P_0/\rho_0} = 1.42 \times 10^{7} \rm{cm \ s^{-1}}$ and the adiabatic soud speed $c_s = \sqrt{\gamma}a_0 = 1.83 \times 10^{7} \rm{cm \ s^{-1}}$. The dimensionless speed of light $\mathbb{C} = c/a_0 = 2.1 \times 10^{3}$ and the ratio of radiation pressure to gas pressure $\mathbb{P} = a_r T_0^4/P_0 = 10^{-3}$.

\section{Results}
\label{sec:results}
\begin{table*}
\begin{center}
    \begin{tabular}{| l | c | c | c| c |c | c | c| c| c|c | c|}
    \hline \hline
	&\multicolumn{4}{c}{Opacity Properties}	& \multicolumn{2}{c}{Timescales \ $[t_{\rm{sc}}]$} &  & \multicolumn{2}{c}{$t = 0.6$} & \multicolumn{2}{c}{$m=2/3m_0$} \\ 
Model	& $\tau_a$ & $\sigma_{a,0}$ & $s$	& $\partial \sigma_a / \partial T$ & $t_{\rm{dif}}$	&$t_{\rm{th}}$  & Summary  & $v$ & $v_{\rm{cm}}$ &  $v$ & $v_{\rm{cm}}$ \\ \hline \hline
tau-1	& $2.0$			& $1.0 \times 10^{-1}$	& $-1$ & $> 0$ & $8.6 \times 10^{-5}$ 	&  $1.4 \times 10^{-3}$  & Rocket & 0.79 & 0.38 & 0.64  & 0.30 \\ \hline
tau0	& $2.0$			& $5.0 \times 10^{-4}$& $0$ & $= 0$ & $8.6 \times 10^{-5}$ 	&  $1.4 \times 10^{-3}$  & Balloon & - & - & - & - \\
tau+1	& $2.0$			& $2.5 \times 10^{-6}$	& $1$ & $< 0$ & $8.6 \times 10^{-5}$ 	&  $1.4 \times 10^{-3}$  & Balloon & - & - & - & - \\ \hline
sigma0	& $4.0 \times 10^2$	& $1.0 \times 10^{-1}$	& $0$ & $= 0$ & $1.7 \times 10^{-2}$ 	&  $6.7 \times 10^{-6}$  & Rocket & 0.58 & 0.54  & 0.6  & 0.41 \\
sigma+1& $8.0 \times 10^4$	& $1.0 \times 10^{-1}$	& $1$ & $< 0$ & $3.4 \times 10^{0}$ 	&  $3.4 \times 10^{-8}$  & Rocket & 0.59 & 0.60 & 0.57   & 0.50 \\ \hline \hline
    \end{tabular}
\end{center}
\caption{Summary of monotonic opacity cloud models. We indicate the cloud optical depth, $\tau_a$, the absorption cross section $\sigma_{a,0}$, the opacity power-law scalings on temperature $s$ (see equation (\ref{eq:absorption})). We list the corresponding diffusion $t_{\rm{dif}}$ and thermal $t_{\rm{th}}$ time scales in units of the sound crossing time $t_{\rm{sc}}$ and a qualitative description of the dynamics. For clouds undergoing rocket acceleration we indicate the cloud core velocity $v$ and center of mass velocity $v_{\rm{cm}}$ of the cold gas at representative time $t = 0.6 \rm{s}$ and when the cloud has evaporated to $m = 2/3 m_0$ of its initial mass.}
\label{tab:summary}
\end{table*}

We investigate models with a variety of opacity coefficients of the form (\ref{eq:absorption}), with power-law scalings ranging from $-1 \leq s \leq 1$ and $n = 2$. We describe the basic physics governing the dynamics of these monotonic power-law models in Section \ref{sec:monotonic}. The behaviour of these models is qualitatively different because as the cloud heats from the incident radiation, the opacity decreases ($s = -1$), remains constant ($s = 0$) or increases ($s = 1$).  

More realistic modeling using photoionization codes find that gas opacity is not a simple power-law but has features due to specific chemical elements, notably H, He and Fe. In Section \ref{sec:turnover} we investigate the effect of such features by studying models where the opacity is a broken power-law, turning over from $s = -1$ to $s = 1$ at a critical temperature $T_c$. As the cloud heats, this turnover in opacity can allow the evaporating gas to become optically thin and potentially avoid the runaway heating that leads to cloud dispersal in our models with simple power-law scaling of opacity.

\subsection{Monotonic Power-law Opacity}
\label{sec:monotonic}
We explore the effects of temperature dependence on opacity for power-law scalings of the form (\ref{eq:absorption}). A summary of our models is shown in Table \ref{tab:summary}, where we indicate the relevant parameters such as cloud optical depth $\tau_a$, absorption cross section $\sigma_{a,0}$, power-law scaling $s$ and the sign of the opacity slope $\partial \sigma_a/ \partial T$. We also list the various time-scales of the model, the diffusion time $t_{\rm{dif}} = 4 x_0^2 \sigma_t/\mathbb{C}$ and the thermal time $t_{\rm{th}}= P/(\mathbb{P}\mathbb{C}E_r\sigma_a)$ in units of the sound crossing time $t_{\rm{sc}} = 2x_0/c_s$. Finally we list the qualitative behaviour of the cloud due to irradiation. For clouds undergoing significant acceleration we list the velocity of the core $v$ and center of mass $v_{\rm{cm}}$ at representaitve time $t = 0.6$ and when the cloud has mass $m = 2/3 m_0$ remaining.  

\begin{figure*}
                \centering
                \includegraphics[width=1.\textwidth]{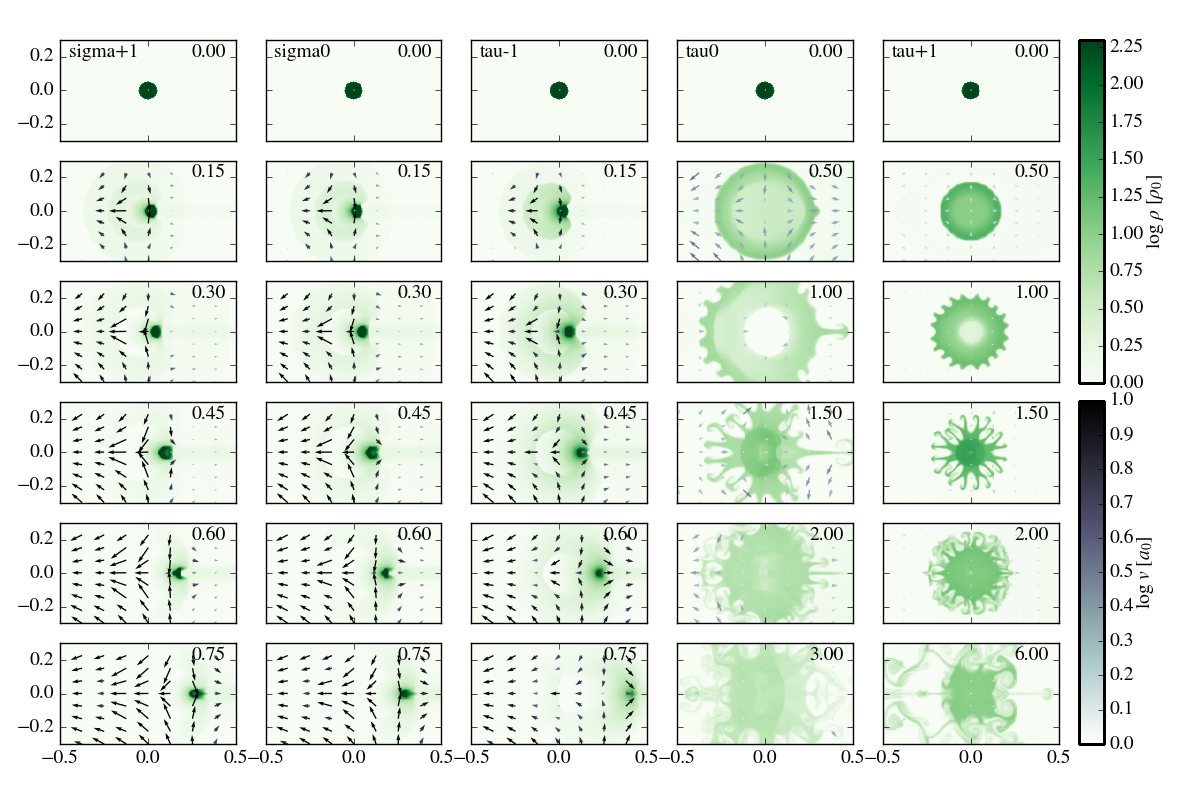}
        \caption{Summary of models with monotonic opacity law showing density (green scale) and velocity vectors (grey scale) in units of the sound speed at representative moments in time in seconds (top right of each panel). Radiation flux enters from the left boundary and interacts with the cloud. The fiducial model with $\sigma_{a,0} = 0.1$ and $\tau_a = 2$ is in the center column. Models to the left have constant initial cloud optical depth $\tau_a$ and models to the right have constant ambient gas opacity $\sigma_{a,0}$. When clouds heat non-uniformally (models to the left) they accelerate via the rocket effect whereas models that heat nearly uniformally (models to the right) cause the cloud to balloon and dissipate.}
\label{fig:density_summary}
\end{figure*}

\begin{figure*}
                \centering

                \includegraphics[width=1.\textwidth]{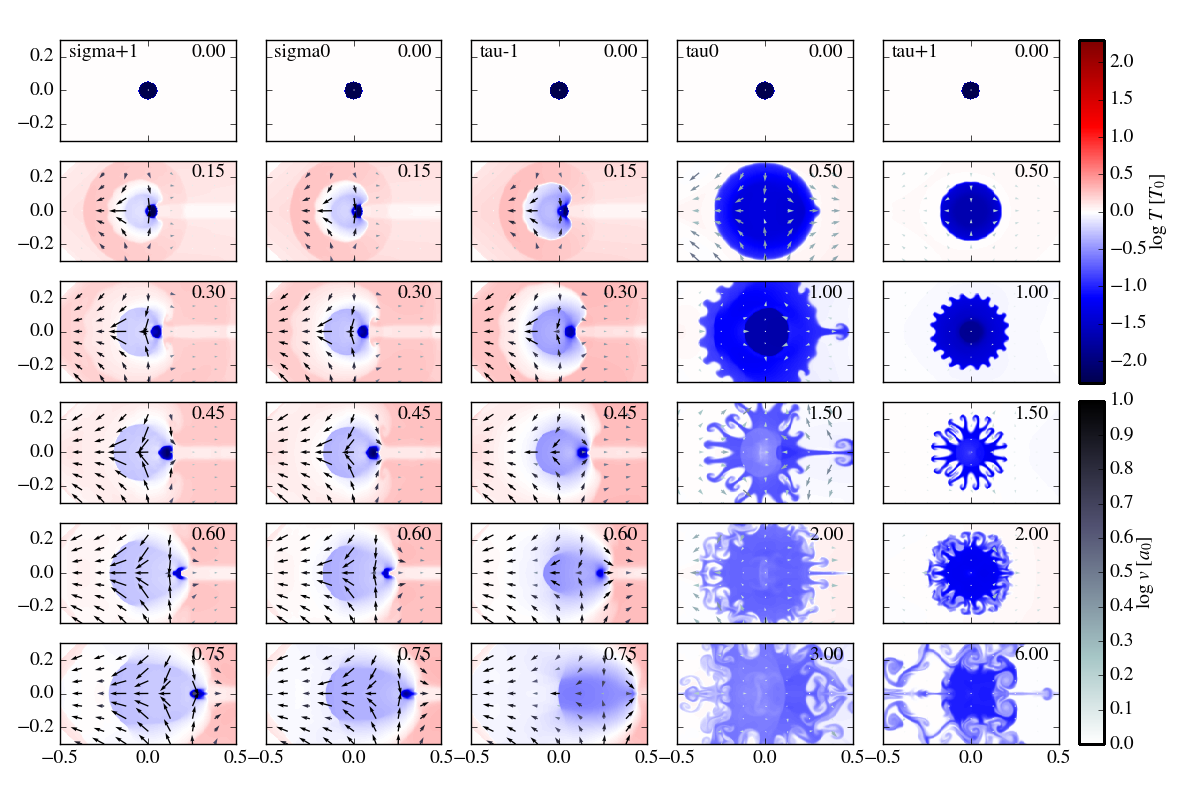}
        \caption{Same as Fig \ref{fig:density_summary} but showing logarithmic temperature contours (color). When clouds heat non-uniformally (models to the left) they accelerate via the rocket effect whereas models that heat nearly uniformally (models to the right) cause the cloud to balloon and dissipate.}
\label{fig:temp_summary}
\end{figure*}

We summarize our results with snapshots of the density (Fig. \ref{fig:density_summary}) and temperature (Fig. \ref{fig:temp_summary}) for models with $-1 \leq s \leq 1$ and either $\tau = \rm{const}$ (three leftmost columns) or $\sigma_{a,0} = \rm{const}$ (three rightmost columns). Our fiducial model (center column) has $s = -1$, an initial cloud optical depth $\tau = 2$ and $\sigma_{a,0} = 0.1$. In this regime the cloud is optically thick for the duration of the simulation, since it is initially optically thick and $s < 0$ ensures that opacity \emph{increases} as it heats.  Clouds exhibit different qualitative behaviour across the parameter space of models. The clouds can accelerate via the \emph{rocket effect} or diffuse away like a \emph{balloon} (see also P14, models A40 and A10 respectively).

\begin{figure*}
  \begin{adjustbox}{addcode={\begin{minipage}{\width}}{\caption{%
      Dynamical variables as a function of time for monotonic power-law opacity models. \textit{Top -} Radiation flux $F_{rx}$ as a function of time. This shows in what cases the cloud is optically thick and the time required for radiation to diffuse through the cloud. \textit{Second from top -} Minimum radiation flux $F_{\rm{rx,min}}$ (solid line) exiting the right side of domain. \textit{Third from top -} Mass of cold gas $M$ (solid line) and maximum cloud density $\rho_{\rm{max}}$ (dashed line) all normalized  to initial cloud mass. \textit{Third from bottom -} Center of mass of cold gas $x_{\rm{cm}}$ (solid line) and position of cloud core $x$ (dashed line). \textit{Second from bottom -} Cold gas center of mass velocity $v_{\rm{cm}}$ (solid line) and velocity of cloud core $v$ (dashed line).  \textit{Bottom -} Kinetic (solid line), thermal (dashed line) and radiation (dotted line) energy density in the cloud. Radiation energy is dominant in all models. Accelerating cloud models have $E_{\rm{th}} \simeq E_{\rm{K}}$ whereas ballooning models have $E_{\rm{th}} \gg E_{\rm{K}}$}\label{fig:dynamical} \end{minipage}},rotate=90,center}
\includegraphics[width=1.3\textwidth]{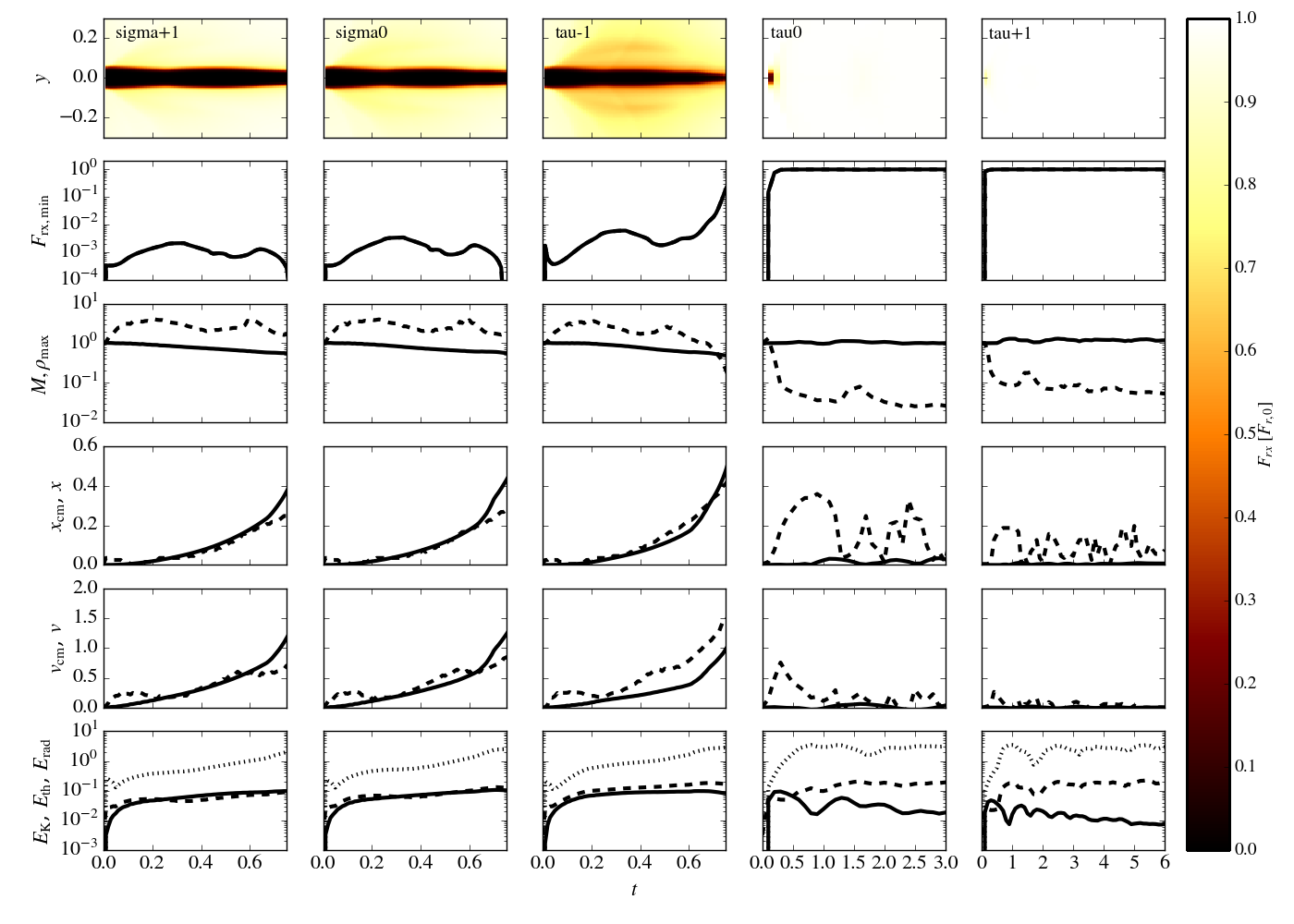}
\end{adjustbox}
\end{figure*}

In Fig \ref{fig:dynamical}, we plot some dynamical variables of the cloud models as a function of time. The top panel shows the radiation flux exiting the far boundary $F_{rx}$, normalized to the flux entering the domain $F_{r,0}$. The second panel shows the minimum radiation flux exiting the far boundary $F_{\rm{rx,min}}$ i.e the radiation passing through the most optically thick part of the cloud. The third panel shows the total cloud mass $M$ (solid line) and the maximum cloud density $\rho_{\rm{max}}$ (dashed line). Here we define gas to be part of the cloud when $T < T_0$ i.e. the gas is cold, relative to the ambient gas. The fourth panel shows the center of mass position of the cloud (i.e the cold gas) $x_{\rm{cm}}$ (solid line) and the position of the cloud core $x$ i.e., position of the density maximum. The second to last row of panels shows $v_{\rm{cm}}$ (solid line) and $v$ (dashed line) in the $\hat{x} $ direction. The bottom panels shows the cloud kinetic energy $E_K$ (solid line), thermal energy $E_{\rm{th}}$ (dashed line) and radiation energy $E_{\rm{rad}}$ (dotted line). 

A useful time scale for this problem is the sound crossing time, $t_{sc}$. Since it is fixed across our models, it is natural to use it to compare the various other time-scales to. 

We see two qualitatively different behaviours, based on the hierarchy of the thermal and diffusion time. When the cloud heats \emph{non-uniformally} it accelerates via the rocket effect. Diffusion through the cloud is slow compared to the rate of heating, $t_{\rm{dif}} \gtrsim t_{\rm{th}}$. The irradiated side heats up and the back-reaction of this evaporating gas accelerates the cloud. When the cloud heats \emph{uniformally} the gas expands outwards like a balloon. Physically this regime requires photons to diffuse through the cloud faster than they can heat it, that is to say $t_{\rm{dif}} \lesssim t_{\rm{th}}$. Since $t_{\rm{th}}/ t_{\rm{dif}} \sim \tau_a^{-2}$, this occurs when the cloud is optically thin. 

Models sigma\_s (three leftmost columns of Fig \ref{fig:density_summary}) are initially optically thick and remain so throughout the simulation, irrespective of the sign of $s$. To become optically thin the cloud density and temperature would have to respectively decrease and increase to the ambient backgrounds values. This is not possible, and therefore these models remain optically thick throughout the entire simulation. Clouds undergo acceleration via the rocket effect and exit the simulation in approximately the sound crossing time. We consider two possible metrics for quantifying the cloud acceleration. The cloud velocity $v$ is defined at the density maxima whereas the center of mass velocity $v_{\rm{cm}}$ is the density weighted velocity over all cold gas, that is to say with $T < T_0$. As figures of merit we list $v$ and $v_{\rm{cm}}$ at $t = 0.6$ in Table \ref{tab:summary}.  We see a slight dependence on the temperature scaling and $v$, with the highest velocity acheived for $s = -1$. In this model, cloud opacity increases as it heats, thereby increasing evaporation rate as it is irradiated by an ever increasing flux (second row of Fig \ref{fig:dynamical}). The higher evaporation rate allows the core to be irradiated by higher levels of flux (due to a decreased column density from the higher evaporation rate) which leads to a higher acceleration and thus larger cloud velocity. At $t=0.6$ for instance $s=-1$ cloud has $v$ $\sim 35 \%$ faster than the other models.  The evaporating gas however is not heated as much as in the other models because of a decreased flux through the cloud atmosphere (top row of Fig \ref{fig:dynamical}). Consequently, much of the evaporating gas with $v < 0$ is not heated above $T_0$ and thus makes a negative contribution to $v_{\rm{cm}}$. This results in the $s=-1$ model having $v_{\rm{cm}}$ $\sim 35 \%$ slower center of mass velocity than the other models. This effect is less pronounced for the $s = 0$ case, resulting in $v_{\rm{cm}}$ $\sim 10\%$ slower than the $s = 1$ case. We can summarize by saying when $s>0$ the evaporating cold gas is more optically thick than for $s \leq 0$ and shields the cloud core from radiation and reduces $v$. However, this evaporating gas is more quickly heated and makes a smaller negative contribution to the center of mass velocity resulting in a higher $v_{\rm{cm}}$ than when $s \leq 0$. Models with $s \geq 0$ thus have a similar $v$ and $v_{\rm{cm}}$, whereas models with $s \leq 0$ efficiently accelerate their core increasing $v$ but conservation of momentum dictates that evaporating cold gas efficiently acquires a large fraction of this velocity thereby reducing $v_{\rm{cm}}$. 

To further quantify the efficiency of the rocket effect for accelerating clouds, we compare velocities after the cloud has evaporated to $m = 2/3 m_0$ of its initial mass. Both the core and center of mass velocities are listed in Table \ref{tab:summary}. By this metric the cloud opacity plays little role in determining the acceleration of the core. The $s=-1$ model is only $\sim 10\%$ higher velocity than model with $s \geq 0$, which is expected from our intuition from the rocket equation as proposed by OS55. The authors modified the standard derivation of the rocket equation as applied to an irradiated cloud and found
\begin{equation}
v = v_e \ln \left( \frac{m}{m_0}\right),
\label{eq:rocket}
\end{equation} 
where $m_0$ is the initial cloud mass and $v_e$ the velocity of the evaporating gas. For a fixed mass loss, the velocity is set by $v_e$. Approximating $v_e \approx c_s$ for the hot gas, when $m = 2/3m_0$ we expect $v \sim 0.4 c_s = 0.52 a_0$, which agrees with our result to within $\sim 20\%$. The energy density in the radiation field is much larger than the kinetic and thermal energy, which are nearly in equipartition. This equipartition is what we would expect from gas evaporated from other objects such as a disc or a star with a thermal wind.

Models tau\_s (two rightmost columns of Fig \ref{fig:density_summary}) are initially optically thick but this changes as the clouds heat since $s \geq 0$. The radiation diffuses through the cloud faster than it causes the cloud to evaporate, so the radiation energy density in the cloud is nearly uniform and the whole cloud heats approximately at the same rate. We can see this as well from the plot of radiation flux exiting the right boundary. Models tau0 and tau+1 have nearly all their flux passing through the cloud a short time after the start of the simulation, whereas tau-1 needs about half the simulation time for the flux to exit. The qualitative behaviour of the cloud thus depends on the optical depth of the cloud.

As radiation saturates the cloud, it heats up, causing the central part to expand outwards (t = 0.5). The accelerating, over dense cloud contacts the stationary, less dense ambient gas producing a shock (t = 1.0). This triggers the Richtmyer-Meshkov instability (RMI, see Brouillette 2002 for a review), producing cold, dense fingers around the cloud. The instability induces a reverse-shock, causing the cloud to recollapse (t = 1.5) after which the cloud slowly dissipates. We see this from the plot of position where the cloud core $x$ undergoes damped harmonic motion but the cold gas center of mass essentially remains at $x_{\rm{cm}} \approx 0$.  After the initial expansion phase generated by the radiation, the evolution of the cloud is primarily driven by pure hydrodynamics. This is characteristic of the (RMI) being a purely hydrodynamic instability, that only requires an accelerating dense medium and is agnostic of the particular acceleration mechanism. We tested this by turning off the radiation field after triggering the initial cloud expansion and found that cloud evolution was largely unchanged. We see a clear hierarchy of scales in the energy, with $E_{\rm{rad}} \gg E_{\rm{th}} \gg E_{\rm{K}}$, consistent with what we expect for a nearly stationary cloud that evaporates as energy is transfered from the radiation field to the gas.

\newpage

\subsection{Broken Power-law Opacity}
\label{sec:turnover}

\begin{table}
\begin{center}
    \begin{tabular}{| l | c | c| c| c | c| c | c|}
    \hline \hline
		\multicolumn{1}{c}{} & \multicolumn{3}{c}{Opacity Properties}	& \multicolumn{2}{c}{$t = 0.6$} & \multicolumn{2}{c}{$m=2m_0/3$} \\
Model		& $T_c \ [T_0]$ & $s_1$ & $s_2$ & $v$	& $v_{\rm{cm}}$	& $v$	& $v_{\rm{cm}}$ \\ \hline \hline
$T_c1/4$	& $1/4$		& $-1$	& $1$ & 1.56	& 0.40		& 1.96 & 0.47 	\\
$T_c1/2$	& $1/2$		& $-1$	& $1$ & 0.85	& 0.56		& 0.98 & 0.65 	\\
$T_c3/4$	& $3/4$		& $-1$	& $1$ & 0.84	& 0.51		& 0.87 & 0.54 	\\ \hline
$T_c\infty$	& $\infty$	& $-1$	& $-$ & 0.79	& 0.38		& 0.64 & 0.30 	\\ \hline \hline
    \end{tabular}
\end{center}
\caption{Summary of broken power law opacity cloud models which all accelerate via the rocket effect. We indicate the critical temperature $T_c$, the power law scalings of the absorption cross section $s_1$ and $s_2$  (see equation (\ref{eq:broken})). We list the cloud core velocity $v$ and center of mass velocity $v_{\rm{cm}}$ at representative time $t = 0.6$ and after the cloud mass has decreased to $m=2/3m_0$ as a metric for acceleration efficiency.}
\label{table:broken}
\end{table}

\begin{figure*}
                \centering
                \includegraphics[width=0.95\textwidth]{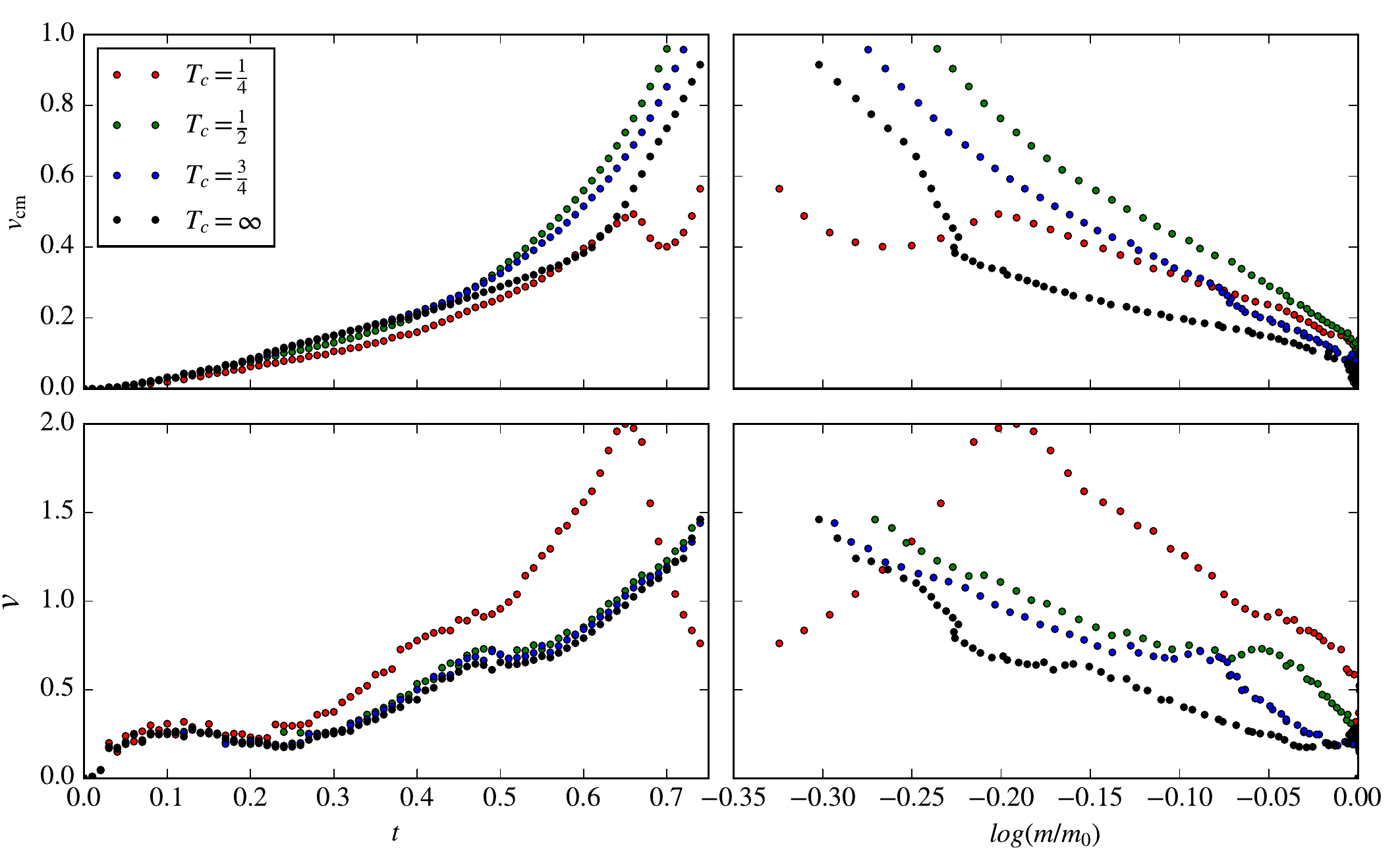}
        \caption{Broken power law models for critical temperatures $T_c = 1/4 \ T_0$ (red), $1/2 \ T_0$ (green) and $3/4 \ T_0$ (blue). We also include the monotonic model tau-1, denoted $T_c = \infty$ (black). Points are plotted in the range $0 \leq t \leq 0.75$ in intervals $\Delta t = 0.01$. Center of mass velocity $v_{\rm{cm}}$ (top panels) and core velocity $v$ (bottom panels) as a function of time (left panels) and cloud mass (right panels). At early times, center of mass velocities are approximately equal but at late times modesl with $1/2 \leq T_c/T_0 \leq 3/4$ are approximately 50\% faster. }
\label{fig:mass_rocket}
\end{figure*} 

We consider models where the absorption cross section is a broken power law,
\begin{equation}
\sigma_a = \begin{cases} 
      \sigma_{a,0} \left( \frac{T}{T_0}\right)^{-s_1} \left( \frac{\rho}{\rho_0}\right)^{n} & T < T_c, \\
      \sigma_{a,0} \left( \frac{T_c}{T_0}\right)^{s_2-s_1} \left( \frac{T}{T_0}\right)^{-s_2} \left( \frac{\rho}{\rho_0}\right)^{n} & T \geq T_c, 
   \end{cases}
\label{eq:broken}
\end{equation}
where $s_1 < 0 < s_2$ so that the cloud opacity initially increases with temperature but \emph{may} avoid runaway heating and dissipating away by turning over at the critical temperature $T_c$.

We set $\tau_a = 2$ and choose $s_1 = -1$ and $s_2 = 1$. The critical temperature is chosen as $1/4 \ T_0 \leq Tc \leq 3/4 \ T_0$. In addition, we can consider the monotonic model tau\_-1 as a limiting case with $T_c = \infty$.  A summary of the broken power-law models are listed in Table \ref{table:broken}, as well as the core and center of mass velocities at representative time $t = 0.6$ and cloud mass $m = 2/3 m_0$. 

All clouds in these models accelerate via the rocket effect as described in the previous section. Models with lower critical temperature have lower temperature evaporating gas, with the atmosphere in case $T_c = 1/4$ most closely resembling the balloon behavior of optically thin monotonic models.  Our goal is to quantify how the acceleration efficiency has been affected by introducing a turnover in the opacity scaling, as would be the case if there is an opacity bump.

In Fig \ref{fig:mass_rocket} we plot $v$ (top panels) and $v_{\rm{cm}}$ (bottom panels) as a function of elapsed time (left panels) and cloud mass (right panels) for different critical temperatures $T_c$. We find that decreasing the critical temperature leads to an increase in the core velocity. For example, at $t=0.6$ the $T_c1/4$ cloud is $85 \%$ faster than the other models. Evaporating gas from the cloud, and thus accelerating it, requires heating gas to $T \sim T_c$. For lower critical temperature the necessary energy deposition required to heat the gas to $T_c$ is lower. Therefore a fixed radiation flux can achieve a higher acceleration. Unlike with monotonic models, the largest cloud velocity is achieved in the case of the most optically thick cloud. In other words, despite the reduced flux incident on the cloud core because of a more optically thick atmosphere, the acceleration is still higher because the cloud atmosphere tends to heat to $T \sim T_c$, and thus have a higher $v_e$, when evaporating. Any additional heating beyond $T \sim T_c$ causes the evaporating gas to become more optically thin and tends to stop subsequent heating of the atmosphere as the coupling between radiation and gas is weakened. The lower $T_c$ cases thus have a denser, colder and hence more optically thick atmosphere (since $s = -1$ for $T < T_c$).

Similarly, after considering the cloud mass as a proxy for acceleration efficiency we find that decreasing the critical temperature increases the core velocity at fixed mass. This is simply a consequence of gas ceasing to heat above $T \sim T_c$. The $T_c1/4$ cloud effectively evaporates gas to accelerate it core, but this gas remains cool $T < T_0$. It therefore remains part of the overall cloud mass budget and gives the impression that this case is much more efficient at accelerating clouds. When we consider the center of mass velocity, we find that the broken power-law models have $v_{\rm{cm}}$ $50 - 100 \%$ faster than the monotonic model $T_c\infty$. We thus conclude that the broken power-law models, by both metrics, are more efficient at accelerating cold gas.

\section{Discussion}
\label{sec:discussion} 

Cloud acceleration can be either momentum or energy driven, depending on which quantity is tranfered from the radiation field to the cloud. Either regime can be achieved in a variety of ways. Momentum can be imparted on the cloud via a pure scattering coefficient, as in P14 (see their models S10 or S200). Alternatively, in Z18 clouds are momentum driven (see their model T1L) in a pure absorption regime since $T_r = T_g$ ensures there is negligible heat transfer. The momentum transfered to the cloud is a function of the incident flux - the optically thick case differs from the optically thin case only by an attenuation factor (see for example equation (25) in Z18). For constant incident flux, as studied in these models, this means that momentum driven acceleration will be approximately constant. Finally, when $T_r > T_g$ clouds can be accelerated via the rocket effect. When clouds are too diffuse or the absorption too strong clouds will disperse before any significant acceleration (P14 case A10) or accelerate reach roughly the sound speed before dispersing (P14 cases A40 or A80). 

We have sought to bridge the regimes explored in previous numerical simulatons of cloud acceleration, by fully considering different temperature dependencies on opacity. We found the RMI is triggered when clouds become optically thin and the rocket effect can be enhanced when clouds become more optically thick. 

The energy driven regime requires a non-zero absorption cross section and a sufficiently massive, optically thick cloud that can absorb hot radiation. P14 studied the case where $\tau \gg 1$, whereas this work considered the cases where $\tau \gtrsim 1$. We can see our clouds are energy driven by comparing the radiative momentum flux incident on clouds and the cloud mass flux, $F_r A/\mathbb{C} \ll \dot{m} v$ where we have $F_r \sim 16$, $A \sim 0.1$, $\dot{m} \sim 3$ and $v \sim 0.5$. Such energy driven winds (see Faucher-Gigu\`ere \& Quataert (2012) and references therein) are supported by observations of AGN that find outflows carry more momentum flux than the radiation field driving them.   

Clouds accelerate via the rocket effect, so velocity increases logarithmically (see equation \ref{eq:rocket} and OS55). If the cloud is insufficiently massive, it dissipates before undergoing significant acceleration as in model A10 of P14. If the cloud is not optically thick, it heats uniformally and expands like a balloon (models tau0 and tau1 of this work). Both momentum and energy driven clouds dissipate as a result of their acceleration. In the momentum driven case, the cloud is shredded by the slower moving, ambient medium. The energy driven cloud dissipates as it is the very gas that is evaporated and responsible for acclerating the cloud. 

We found that optically thin clouds heat uniformally, which causes them to expand and trigger the RMI. Because of our initial mass distribution, this occurs in a circular geometry. Z18 found that clouds are unstable to a Rayleigh-Taylor type instability, which they resolved in their highest resolution runs (T0.01L) with reduced gas pressure. Likewise they found clouds to be Kelvin-Helmholtz unstable in the case of a hot ambient gas ``shredding" clouds (see also Klein et al. 1994; Poludnenko et al. 2002). These findings generally support the conclusion that clouds have a variety of mechanisms by which to dissipate and therefore have a finite lifetime.    

We investigated whether the presence of a ``bump" in the opacity could alter cloud lifetime. In order for a cloud to change from the accelerating rocket regime to the expanding balloon regime the optical depth of the cloud must transition from optically thick to thin. Consider a cloud that is marginally optically thick, $\tau_i \gtrsim 1$ which heats isobarically and has $s_1 < 0$. If the temperature doubles $T_i \rightarrow T_c = 2T_i$, then the opacity changes by a factor $2^{n+s_1}$. If the opacity is a powerlaw, and $s_1 \geq 0$ then we find the cloud evolves in the balloon regime. Suppose that at temperature $T_c$, the opacity function has a break as in (\ref{eq:broken}). A similar doubling of the temperature, $T \rightarrow 2 T_c$ will decrease opacity by a factor $2^{n+s_2}$. For the cloud to re-enter the optically thick regime, we need $s_2 \gtrsim 2n + |s_1|$. With our choice of parameters, we would need $s_2 = 5$, a far steeper power law than say Krammers $s = 3.5$. The steepness of the power-law can be reduced by having a greater temperature change as the cloud thickens, but this range is limited by the initial cloud/medium density contrast. We conclude that except for perhaps some finely chosen area of parameter space it is challenging to change from the optically thin balloon regime to the optically thick rocket regime. We note however than when the transition temperature is low, say $T_c = 1/4$ we did see ballooning of the expanding atmosphere, though the cloud core still underwent rocket acceleration.

\begin{figure}
                \centering
                \includegraphics[width=0.48\textwidth]{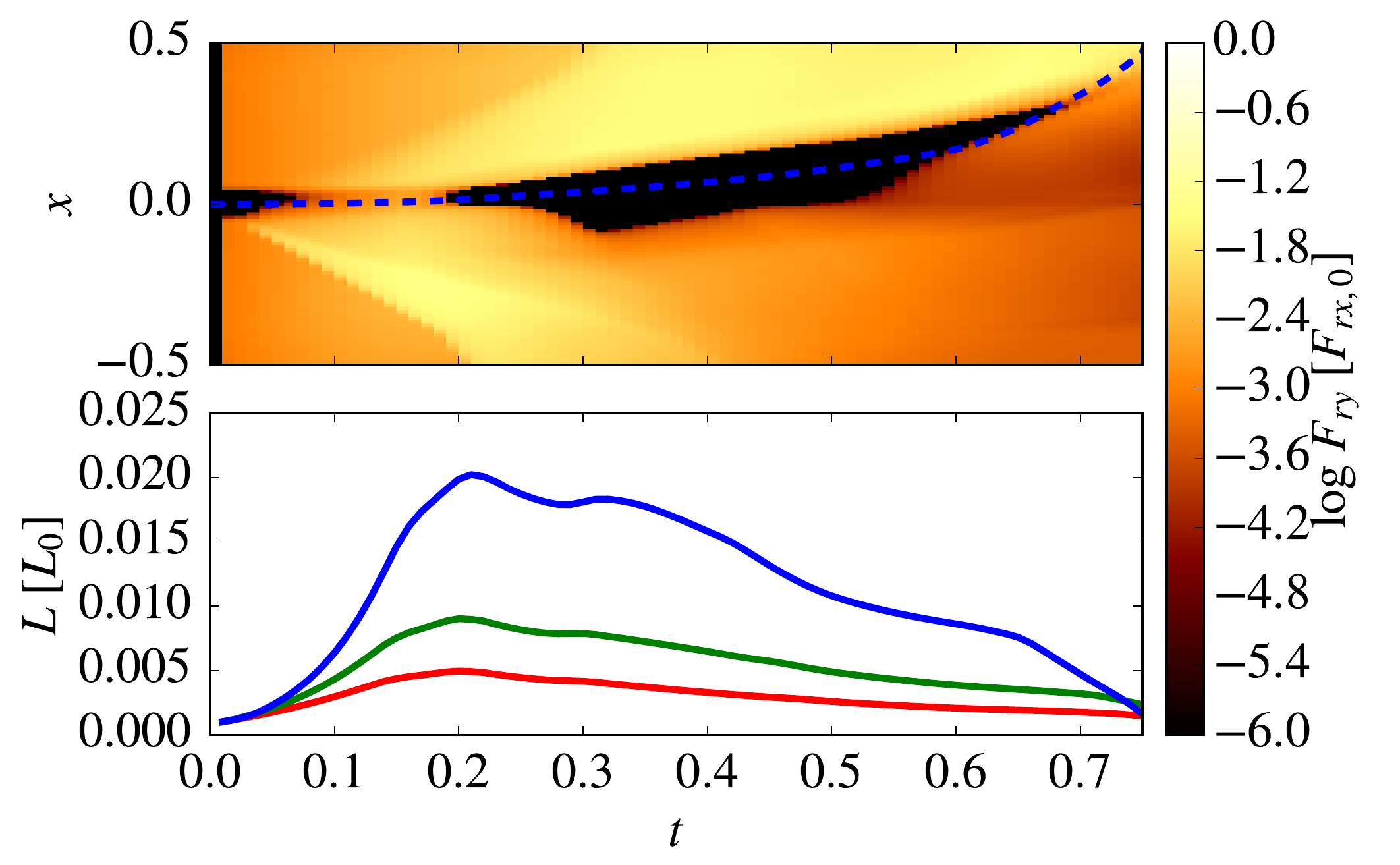}
        \caption{Radiation exiting the domain perpendicular to the cloud motion for sigma1 (red), sigma0 (green) and tau-1 (blue). We do not plot ballooning cloud cases because the outgoing flux is negligible $\lesssim 10^{-5}$ incident flux. \textit{Top -} Radiation flux out the top of the box as a function of time, normalized by the flux incident from the left side, for the tau-1 model. The dashed line indicates the location of the cloud center of mass. The space-time distribution of flux is similar in all cases of an accelerating cloud.  \textit{Bottom -}  Total luminosity exiting the top of the box as a function of time. }
\label{fig:radiation}
\end{figure} 

In Fig \ref{fig:radiation} we plot the radiation exiting the simulation through the top of the box. The top panel shows the spatial distribution of radiation as a function of time (color contour) for the case tau-1. We superimpose the center of mass $x_{\rm{cm}}$ position (blue dashed line).  We see that approximately 10\% of the incident flux is reprocessed by the cloud and re-emitted perpendicular to the incident flux. The core of the cloud itself is not emitting, as we see the low emission regions are well tracked by the center of mass. The bottom panel shows the total luminosity exiting the top of the box, normalized to the incident luminosity for tau-1 (blue), tau0 (green) and tau+1 (red). In all cases the reprocessed emission is $\sim$ few \%. This is what we expect from purely geometric considerations, since this cloud has a covering fraction of $10\%$, so we expect $\sim 1/4$ of this radiation to be re-radiated out the top part of the box or $L \sim 2.5\% L_0$. It peaks at $\sim 0.2$ s, which corresponds to the time when the initial transient phase of gas evaporation from the cloud occurs. It then decreases roughly linearly with time before dropping to zero as the cloud exits the simulation domain. This flux is purely reprocessed emission, since the incident flux was perpendicular to it. We can estimate that the cloud is reprocessing $\sim 1\%$ of the incident radiation, and re-emitting it isotropically. We can thus estimate the line emission and absorption from a single cloud to be approximately 1\%. In the case of an optically thin cloud, the emission is on the order of $\sim 10^{-5}$, even for the case with highest opacity (sigma0). This is consistent with what we expect since as the cloud heats to $T_0$ the optical depth decreases by $\sim 10^{3}$, so we expect the amount of reprocessed radiation  to drop by a similar factor. 

The broken power-law models show a nearly factor $\sim 10$ drop in re-emitted flux. As seen from the spatial distribution of re-emitted flux, most radiation is coming from the hot, evaporating gas. Introducing a cutoff $T_c$ means that this hot gas is less optically thin and therefore reprocesses less incident radiation.

\section{Conclusion}
\label{sec:conclusion}
We have studied the dynamics of a single cloud absorbing radiation from a distant source. We find the cloud behaves in two qualitatively different ways. If the cloud is optically thin, it heats nearly uniformally and expands like a balloon. If the density of the heated gas is higher than of the ambient gas, this triggers the Richtmyer-Meshkov instability and leads to cloud dissipation. If the cloud is optically thick, it heats preferentially on the radiated side and gas evaporation accelerates the cloud via the rocket effect. The velocity growth is logarithmic, quantitatively different from the linear growth seen in the regime where the radiation and gas are in thermal equilibrium as studied in Z18. 

We could not qualitatively alter the behaviour of clouds using broken power-law opacities - accelerating clouds could not be made to balloon and ballooning clouds could not be made to accelerate. We estimate such a qualitative change in behaviour would require a very steep, $s \sim 5$ opacity temperature dependence. However the efficiency of cloud acceleration can be increased if hot gas is more optically thin, as gas evaporating from the cloud no longer absorbs incident radiation. These results suggest that features in the opacity profiles due to certain chemical elements can affect cloud dynamics.  

After having modeled the dynamics of a single cloud, we are in a position to simulate multiple clouds in a dynamic environment. Proga and Waters (2015, see also Waters and Proga 2019) have shown how clouds can form via thermal instability from initial perturbations. Using this initial setup we can form clouds \emph{in situ} and study their evolution in a periodic box. We expect clouds to dissipate as they accelerate/balloon away and reform again via thermal instability. One possible scenario is finding a quasi-steady, multi-phase solution as predicted by Krolik McKee \& Tarter (1981). We may then characterize the covering fraction of such a system, as measured by observations of AGN tori.

We find that with the rocket effect it is difficult to accelerate clouds much beyond the sound speed $v \gtrsim c_s$. However, as we have shown features in the opacity profile can effectively make the evaporating gas optically thin, which allows it to stay cool relative to the ambient gas. If this cool gas, reforms clouds via thermal instability, it may be possible to accelerate cold gas beyond the sound speed after multiple cycles of acceleration and condensation.

We have assumed clouds are initially in hydrostatic pressure balance. Around AGN, radiation pressure from the central object is expected to be important is accelrating clouds, so an equally compelling initial condition is one where the cloud is in pressure equilibrium between its internal gas pressure and the external radiation field (see for example Dopita et al. 2002). A fundamental question is to explore how the dynamics of these clouds may be different from those initially in thermal pressure equilibrium. This is particularly an important question for AGN, where radiation pressure is expected to be an important wind driving mechanism.

\section*{Acknowledgements}
All simulations were performed on the UNLV National Supercomputing Institute's Cherry Creek cluster and the authors acknowledge Ron Young's technical expertise. S.D. acknowledges support from ERC Advanced Grant 340442. C.S.R. thanks the UK Science and Technology Facilities Council (STFC) for support under the New Applicant grant ST/R000867/1, and the European Research Council (ERC) for support under the European Union's Horizon 2020 research and innovation programme (grant 834203). Support for Program number HST- AR-14579.001-A was provided by NASA through a grant from the Space Telescope Science Institute, which is operated by the Association of Universities for Research in Astronomy, Incorporated, under NASA contract NAS5- 26555.  This work also was supported by NASA under ATP grant 80NSSC18K1011.

\label{lastpage}

\end{document}